
\input j-phys.sty

cond-mat/9211020

\title{Diffusion and spectral dimension on Eden tree}

\author{Hisao Nakanishi\footnote\S{Permanent address:
Department of Physics, Purdue University, W. Lafayette, IN
47907 U.S.A.} and Hans J. Herrmann}

\address{HLRZ, KFA--J\"{u}lich, W--5170 J\"{u}lich, Germany}

\shorttitle{Diffusion on Eden Trees}


\jnl{\JPA}

\beginabstract
We calculate the eigenspectrum of random walks on the Eden tree
in two and three dimensions.  From this, we calculate the
spectral dimension $d_s$ and the walk dimension $d_w$ and test
the scaling relation $d_s = 2d_f/d_w$ ($=2d/d_w$ for an Eden tree).
Finite-size induced crossovers are observed, whereby the system
crosses over from a short-time regime where this relation is
violated (particularly in two dimensions) to a long-time regime where
the behavior appears to be complicated and dependent on dimension
even qualitatively.
\endabstract

\section{Introduction}
Treelike structures arise in many situations in statistical physics,
including clusters formed by irreversible growth processes modeled,
e.g., by diffusion-limited aggregation [1] (which are treelike
on large scales) and dielectric breakdown [2].
In this paper, we consider a particularly simple tree structure called
Eden tree [3] which is formed on a lattice by a modification of the
usual Eden process [4] (in which a {\it compact} cluster grows)
so that an empty site which is a neighbor to more than one
occupied site becomes ineligible for occupation.

The study of diffusion on treelike structures is
interesting in that it may present qualitatively different behavior
compared with the structures with many loops (particularly those
that connect large branches).  The Eden tree model was studied in detail
by Dhar and Ramaswamy [3] using various methods.  They calculated
the spectral dimension $d_s$ [5] for two- and three-dimensional
Eden trees by applying the {\it node counting theorem} [6] and also the
random walk dimension $d_w$ [7] by a direct simulation of random walks
as well as through the {\it backbone} statistics of the tree,
and discussed them in terms of a scaling argument.  A surprising
conclusion was drawn from these discussions; they concluded that
a typical long walk samples only order-one segments of the backbone
(including the {\it dangling ends} attached to this segment) so that
the usual scaling relation [8]
$$
d_s = 2 d_f /d_w  \eqno(1)
$$
does not hold. (Here $d_f =d$ since the Eden trees are compact.)
While the calculation of $d_s$ (particularly in two
dimensions) seemed accurate, the graphs from which $d_w$ was deduced
appeared to show considerable curvature.  Thus a more accurate
determination of these exponents appeared
desirable to test the violation of relation (1) better.

The scaling law (1) is expected to be valid in general because it
follows by assuming only: (a) the equivalence of the vibrational and
diffusional problems (Alexander and Orbach [8]) and (b) that a random
walk samples essentially all sites more or less uniformly.  (In this
paper, we shall loosely call the latter condition {\it ergodic}
for simplicity.) In fact this relation
is sometimes used as the {\it definition} of the spectral dimension
$d_s$.  In their work [3], Dhar and Ramaswamy essentially reject
assumption (b) thereby rejecting the relation for the Eden tree.

This is surprising because it appears at first that any random walk
satisfying the detailed balance condition should in principle be
{\it ergodic} (in the way we use the term in this paper).
This should be true independent of the structure
and thus for example (1) holds for Euclidean lattices in all dimensions
and also for critical percolation clusters [9].

In this paper, we reexamine this rather curious observation
of Ref.[3], using a method which allows a more accurate
numerical evaluation of both $d_s$ and $d_w$ from the same data.
This is the method of calculating the eigenspectrum of the so-called
hopping probability matrix ${\bf W}$ by Saad's diagonalization method
[10] and making use of the Laplace transform relationship between
this spectrum and quantities such as the return-to-origin probability
and the velocity autocorrelation function [11,12].  (Here ${\bf W}$
has elements $W_{ij}$ which are just the probability for a random
walker at site $j$ to hop to site $i$ in the next time step.)
This corresponds to averaging over {\it all} random walks starting
from every point of the cluster.
The method was used previously [12,13] for two- and three-dimensional
percolation clusters very successfully and
we expect it to work just as well for the Eden trees.

Thus we first construct an Eden tree of size $S$ by
simulation on the square and simple cubic lattices,
create the matrix ${\bf W}$ for a random walk (we use a {\it blind ant}
model where the random walker has an equal probability to hop between
any pair of neighboring sites),
and calculate the eigenvalues and eigenvectors of ${\bf W}$ for the
largest $M$ (typically a few hundred) eigenvalues to very high
precision (typically up to 6 decimal places)
using Arnoldi-Saad algorithm [10].
The eigenvalue density $n(\lambda )$ is then expected [13] to scale as:
$$
n(\lambda ) \sim | \ln \lambda |^{d_s /2-1} . \eqno(2)
$$
One can also construct another interesting function
$\pi (\lambda )=n(\lambda )a_{\lambda}(\lambda -1)^2$ [14]
from the eigenvalues and eigenvectors of ${\bf W}$,
where $a_{\lambda}$ are the coefficients entering
the mean position autocorrelation
$<{\bf r}(t) \cdot {\bf r}(0)> = \sum_{\lambda} a_{\lambda}{\lambda}^t$.
The coefficients $a_{\lambda}$ are determined when the stationary
initial distribution is expanded in terms of the eigenvectors of
${\bf W}$.  This function is expected [12,13] to scale as
$$
\pi (\lambda ) \sim | \ln \lambda |^{1-2/d_w}  .  \eqno(3)
$$

\section{Numerical results}
Our numerical results for $n(\lambda )$ and $\pi (\lambda )$ are
plotted in Fig.1 and Fig.2 respectively. (The values of $\pi$ plotted
have been scaled up by a factor of the size $S$ of the tree compared
with the convention used in Ref.[13] in order to separate different $S$
in the figure.) The cluster sizes for which the calculations were
made are $S=2500$, $5000$, $10000$ for the square lattice for which $200$,
$400$, and $200$ independent realizations were averaged over
(respectively), and $S=2500$, $5000$ for the simple cubic lattice
for which $200$ and $400$ independent realizations were averaged
(respectively).

{}From Fig.1(a) for the square lattice,
there is apparently a sharp cluster-size dependent
crossover which divides a region of larger $|\ln\lambda |$
(corresponding to shorter time) where
an excellent power-law fit can be made, from the smaller $|\ln\lambda |$
region where another power-law fit may also be possible.
We will see later that the locations of these crossovers are
consistent with the finite size scaling with the walk dimension.
Fig.1(b) for the simple cubic lattice gives
a much less clear indication of such a crossover, but nonetheless,
the sudden flattening of the data in the regions of smallest
$|\ln\lambda |$ is consistent with a crossover interpretation.
We observe that the right-hand side regions in these figures
should give the asymptotic exponent corresponding to Eq.(2)
as $S$ increases (because this is the region before the finite
size effects appear to set in). However, the apparent power-laws
in the left-hand side region are still ununderstood.  Moreover,
it is even possible that the qualitative behavior of these crossovers
changes when $S$ is made much larger; since in the present analysis
we are limited to relatively small Eden trees, we cannot be sure
if there is not another true asymptotic behavior.

Estimating the slopes of the log-log plots
in the right-hand side regions and using Eq.(2),
we obtain
$$
d_s = 1.22 \pm 0.02 \;\;\; (d=2)  \eqno(4)
$$
and
$$
d_s = 1.32 \pm 0.02 \;\;\; (d=3),  \eqno(5)
$$
where the errors quoted are mainly the fitting errors but also include
some error due to the choice of the region to fit and that due to
the small variation for different cluster sizes.
These estimates are consistent with those of Ref.[3] but
improves the accuracy especially in three dimensions.
In particular, we rule out the possibility
that $d_s$ is the same for two and three dimensions.

The results for $\pi (\lambda )$ in Fig.2 show an even more
interesting cluster-size dependent anomaly.  In Fig.2(a) for the
square lattice, starting from the larger
$|\ln \lambda |$ (i.e., shorter time)
region, there is a range of about one to two decades where an excellent
power law fit can be made, and then a sudden and large
decrease occurs.
This anomaly is reproducible in independent batches
of clusters and thus not caused by simple statistical fluctuations.
In Fig.2(b) for the simple cubic lattice, another sharp, but
very different type of crossover is observed, where the small
$|\ln\lambda |$ (or long time) region also gives a power-law
but with a different exponent.

Let us first discuss the results shown in Fig.2(a) for the
square lattice in more detail.  Here, if only the flat region for
larger $|\ln \lambda |$ is fitted to a power-law, the exponent
comes out to be about $0.29\pm 0.01$, $0.29\pm 0.01$,
and $0.19\pm 0.01$ for $S=2500$, $5000$, and $10000$, respectively.
We note that the last value would yield $d_w =2.47\pm 0.03$,
which would be barely consistent with the value $d_w = 2.54 \pm 0.04$
obtained in Ref.[3] from the backbone statistics.  This also means
that these data would violate (1) very strongly, since $2d/d_w$ would
be about $1.62\pm 0.02$ using this value of $d_w$.  Moreover, since
the crossover point where the drop begins moves to smaller
$|\ln \lambda |$ for larger cluster size $S$, we would expect this
exponent to be the {\it correct} asymptotic exponent $d_w$ for
$S \rightarrow \infty$.

Our interpretation of this result is as follows: for sufficiently
small time (thus length) scales ($t << \tau$)
a typical random walk is trapped (by lack of loops) within
order-one segment of the backbone (together with its dangling
side branches) as proposed in [3], which results in a small $d_w$
and the violation of (1).  The crossover time $\tau$ should be
such that
$$
\tau^{1/d_w} \propto L \eqno(6)
$$
where $L$ is the length scale of the cluster and $d_w$ is the
walk dimension in this regime.  Translated to the corresponding
value in $|\ln \lambda |$, the crossover point should scale as
$$
|\ln \lambda_1 | \propto {\tau}^{-1} \propto L^{-d_w} . \eqno(7)
$$

However, as the walk goes beyond
this time scale, it is forced by finite size effect
to sample more and more segments of the backbone.
Thus over long times, an effective $d_w$ increases,
leading to the downward tendency in Fig.2(a) beyond the crossover.
Of course, for even longer times, the size of the visited region
becomes limited by the total size of the cluster and the walk visits
every site of the cluster; however, such a long-time regime is
not reached in our figure.

Numerically, the crossover values $|\ln \lambda_1 |$for $n(\lambda )$
are approximately $5 \times 10^{-4}$, $2 \times 10^{-4}$, and
$1 \times 10^{-4}$ for $S=2500$, $5000$, and $10000$, respectively.
In comparison, $L^{-d_w}$ with $L=\sqrt{S}/2$ and
the estimate $d_w =2.47$ gives
$3.5\times 10^{-4}$, $1.5\times 10^{-4}$, and $6.4\times 10^{-5}$,
in fair agreement with the discussion above.  The
discrepancy of about a factor of $1.5$ in the absolute numbers
should be due to the fact that our method is equivalent to
summing over all random walks starting from all sites of the
cluster and not just from the seed site at the center.

Having identified $d_w$, we can return to the crossovers observed
in Fig.1(a) for $n(\lambda )$ in two dimensions.
These crossovers occur roughly around $|\ln\lambda |=1\times 10^{-4}$,
$5\times 10^{-5}$, and $2\times 10^{-5}$ for $S=2500$, $5000$, and
$10000$, respectively.  The ratios of these numbers are again in fair
agreement with those from Eq.(7).

The situation for $\pi (\lambda )$ in three dimensions is much less
clear.  In Fig.2(b), $\pi (\lambda )$ is plotted for $S=2500$
and $5000$ for the simple cubic lattice.  Similarly to the square
lattice, there is a flat region for larger $|\ln \lambda |$ which
can be fitted to a  power-law and what seems to be a crossover
to another power-law at smaller $|\ln \lambda |$.  The first
region gives slopes of $0.49\pm 0.01$ and $0.51\pm 0.01$
for $S=2500$ and $5000$, respectively.  This would translate to
$d_w$ of about $3.92\pm 0.08$ and $4.08\pm 0.09$, respectively,
and thus, with $d_f =d=3$, we would have
$2 d_f /d_w$ of around $1.5$ which would yield a much smaller
deviation from Eq.(1) than in two dimensions.

Beyond the crossover in the smaller $|\ln \lambda |$ region,
we have another power-law with an exponent of about $0.32$ and $0.29$
for $S=2500$ and $5000$, respectively.  Since these exponents would
lead to smaller values of $d_w$ in this region, we would have an even
stronger violation of the scaling relation (1).

The crossover point is at about $|\ln \lambda_1 |=1\times 10^{-3}$
and $4\times 10^{-4}$ for $S=2500$ and $5000$, respectively.
The ratio of these two numbers is in good agreement with Eq.(7)
when a value of $d_w = 4.08$ is used.  Thus, we tentatively
conclude that the asymptotic behavior for an infinite Eden tree
in three dimensions may also be nonergodic where Eq.(1) is violated
(but less strongly than in two dimensions).  However, we do not
understand the regime for $t >> \tau$. One speculation would
be that, in the latter regime, the random walk does not become
ergodic over the whole cluster but becomes trapped in some
parts of it, e.g., in the surface of the Eden tree. The walk dimension
of this trapping region would then determine the slope after the
crossover.  However, the surface of the Eden model is a particularly
difficult problem from the numerical point of view, with a very slow
and non-monotonic convergence to the asymptotic behavior [15].
It is also possible that the cluster sizes considered here are
not yet in the asymptotic regime and that for much larger clusters
a different picture emerges.  Unfortunately no suitable technique
is available to calculate $d_s$ and $d_w$ accurately on very
large clusters.

\section{Conclusion and discussion}
In this work we have calculated the eigenspectrum of the random walk
hopping probability matrix ${\bf W}$ and from this computed
the critical exponents $d_s$ and $d_w$ for the Eden tree in two
and three dimensions.  By more accurate calculations, we confirm
the observation of Ref.[3] that the random walk is not {\it ergodic}
on two dimensional Eden tree and moreover our data suggest that the same
is true for three dimensions (although less strongly).  Here, we use the
term {\it ergodic} to express the notion that a typical random walk visits
{\it essentially} all sites of the cluster (i.e., at least sites that
have the same fractal dimension as the entire cluster)
in the long time limit.

A natural question is why
the detailed balance in this problem does not guarantee ergodicity.
This is understood when one makes a distinction between {\it ergodicity}
as we used the term here and a different statement that any
initial distribution eventually evolves into the stationary one.
The latter statement is true when detailed balance is satisfied,
but this is a statement about the distribution and {\it not} about
an individual random walk.  Thus it is {\it not} necessary for an
individual random walk to sample {\it essentially} all sites of the cluster
{\it with probability one} even in the $t \rightarrow \infty$ limit.
Rather it can so happen that a typical walk is very anisotropic,
sampling only order-one segment of the tree, but when we consider all
possible walks, the entire cluster is sampled, if one waits long enough.

It is also interesting to consider whether a similar situation arises
in tree structures in general.  A typical treelike structure
encountered in statistical physics is the diffusion limited aggregate
(DLA) [1].  The available numerical results [16] from direct simulation
of random walk displacements indicate that $d_w = 2.56 \pm 0.10$ in $d=2$
and $3.33 \pm 0.25$ in $d=3$, while the simulation of the return-to-origin
probability in the same reference gives $d_s = 1.20 \pm 0.1$ in $d=2$ and
$1.30 \pm 0.1$ in $d=3$. Using the known estimates [5] of
$d_f = 1.68 \pm 0.05$ in $d=2$ and $2.5 \pm 0.06$ in $d=3$ together
with the estimates for $d_w$ from [16], we obtain
the value of $2d_f /d_w$ to be about $1.31 \pm 0.1$ in $d=2$
and $1.50 \pm 0.1$ in $d=3$. While these values are not inconsistent
with the direct estimates of $d_s$ from [16], they are sufficiently
different (particularly in $d=3$) to warrant further investigation.

Indeed, it may be possible that more or less all treelike structures
force a typical random walk to be anisotropic. If this were the case,
it would be of great interest to develop a systematic theory of
this anisotropy, perhaps with different coherence length exponents
characterizing the radial and tangential directions in a manner
somewhat similar to directed percolation [9].  This is, however,
clearly out of scope for this work and must await further research.

\ack
We thank M. Sahimi for discussions.  Also,
H. N. is grateful to HLRZ (German Supercomputing Center) at
KFA--J\"{u}lich for the kind hospitality.

\references

\numrefjl{1}{Witten T A and Sander L M 1983}{\PR B}{\bf 27}{5686--5697}
\numrefjl{2}{Niemeyer L, Pietronero L and Wiesmann H J 1984}{\PRL}{\bf
52}{1033--1036}
\numrefjl{3}{Dhar D and Ramaswamy R 1985}{\PRL}{\bf 54}{1346--1349}
\numrefbk{4}{Eden M 1961}{Proceedings of the Fourth Berkeley Symposium in
Mathematics, Statistics and Probability}{ ed. J. Neyman (Berkeley: Univ. of
California) p.233}
\numrefjl{5}{Havlin S and Ben-Avraham D 1987}{Adv. Phys.}{\bf 36}{695}
\numrefbk{6}{Hori J 1968}{Spectral Properties of Disordered Chains and
Lattices}{(Oxford: Pergamon)}
\numrefjl{7}{Gefen Y, Aharony A and Alexander S 1983}{\PRL}{\bf 50}{77}
\numrefjl{8}{Alexander S and Orbach R 1982}{J. de Phys. Lett.}{\bf 43}{625}
\numrefbk{9}{Stauffer D and Aharony A 1992}{Introduction to Percolation
Theory}{(London: Taylor and Francis)}
\numrefjl{10}{Saad Y 1980}{Linear Algebra Appl.}{\bf 34}{269}
\numrefjl{11}{Harris A B, Meir Y and Aharony A 1987}{\bf 36}{8752}
\numrefjl{12}{Fuchs N H and Nakanishi H 1991}{\PR A}{\bf 43}{1721}
\numrefjl{13}{Nakanishi H, Muckherjee S and Fuchs N H 1992}{preprint}{}{}
\numrefjl{14}{Jacobs D J and Nakanishi H 1990}{\PR A}{\bf 41}{706}
\numrefbk{15}{Herrmann H J 1991}{The Monte Carlo Method in Condensed Matter
Physics}{ed. K. Binder (Berlin: Springer-Verlag)}
\numrefjl{16}{Meakin P and Stanley H E 1983}{\PRL}{\bf 51}{1457}

\figures

\figcaption{(a) The eigenvalue density $n(\lambda )$ for the Eden
tree on the square lattice.  The data points $\bigcirc$, $\times$, and
$\Delta$ correspond to the average over $200$, $400$, and $200$
clusters of size $S=2500$, $5000$ and $10000$, respectively.
The solid lines are obtained by least squares fitting to the data
for $|\ln \lambda |$ larger than its crossover value (number of
points used in the fits being $14$, $19$, and $19$ for the three sizes
respectively), and the dotted lines are from the similar fits
to the data smaller than the crossover point;
(b) $n(\lambda )$ for the simple cubic lattice where the symbols
$\bigcirc$ and $\times$ correspond also to $S=2500$ and $5000$,
respectively.
}

\figcaption{(a) The function $\pi (\lambda )$ for the Eden tree
on the square lattice.  The symbols $\bigcirc$, $\times$, and $\Delta$
have the same meaning as in Fig.1;
(b) Same for the simple cubic lattice.
}

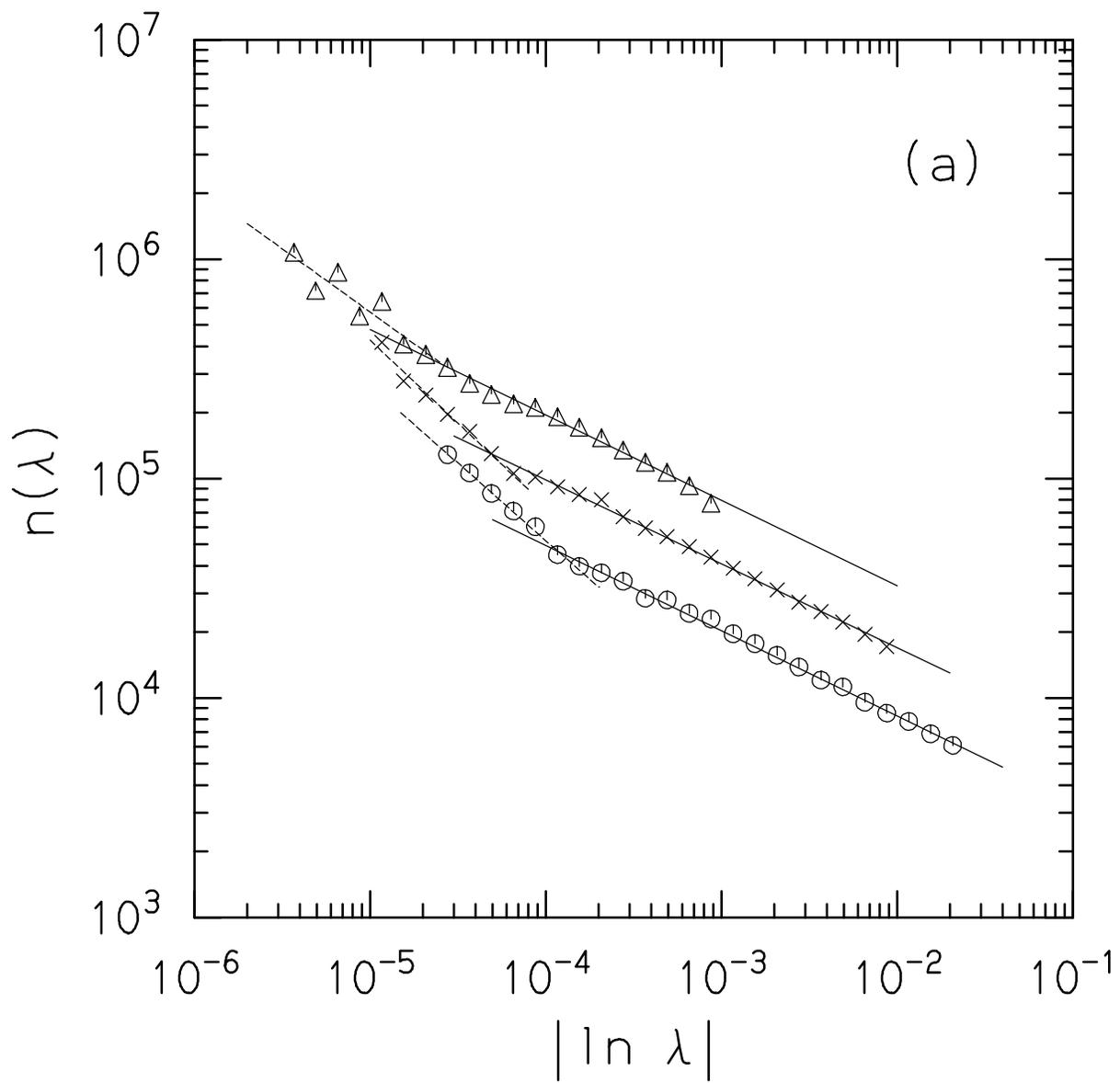
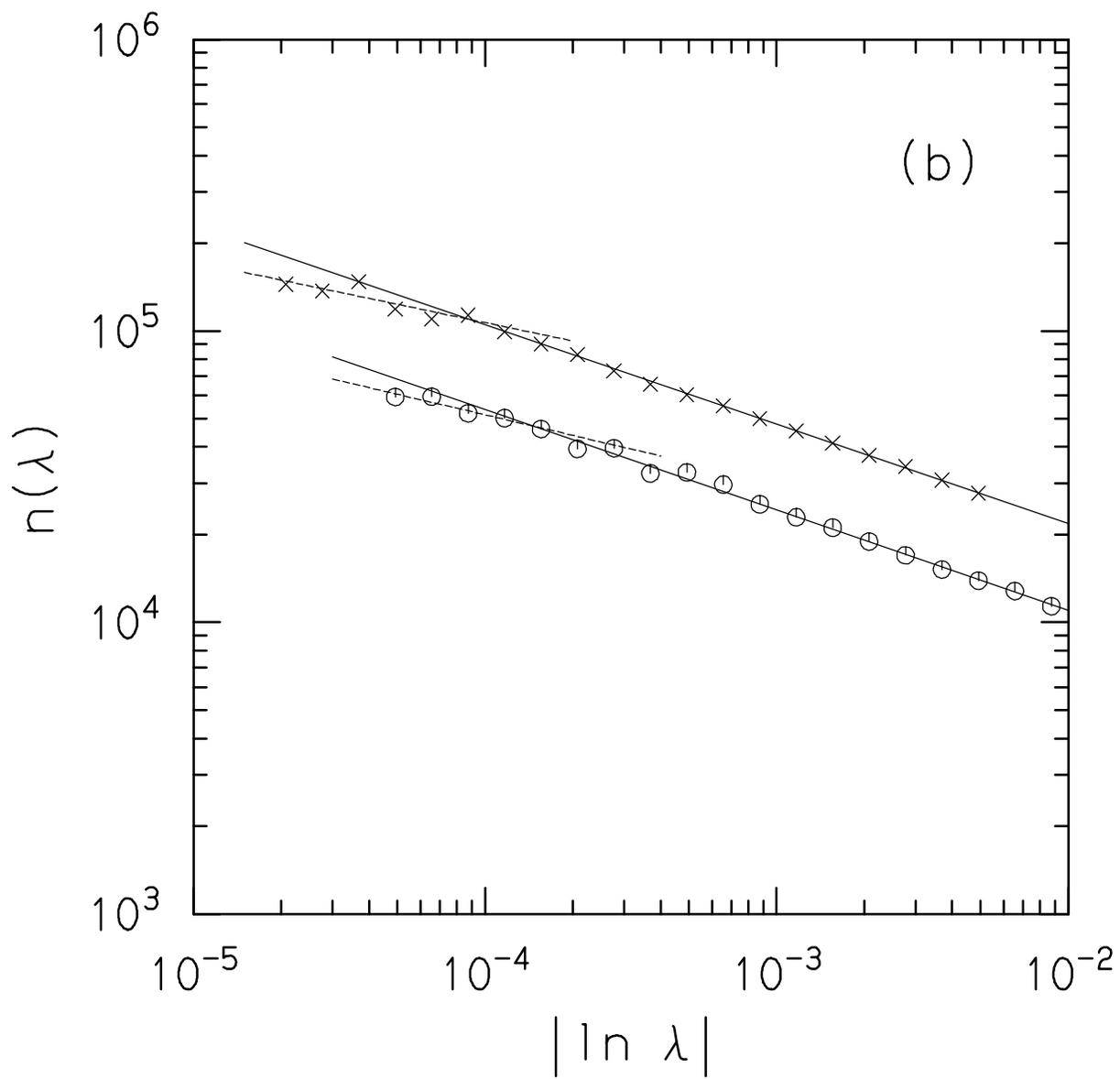
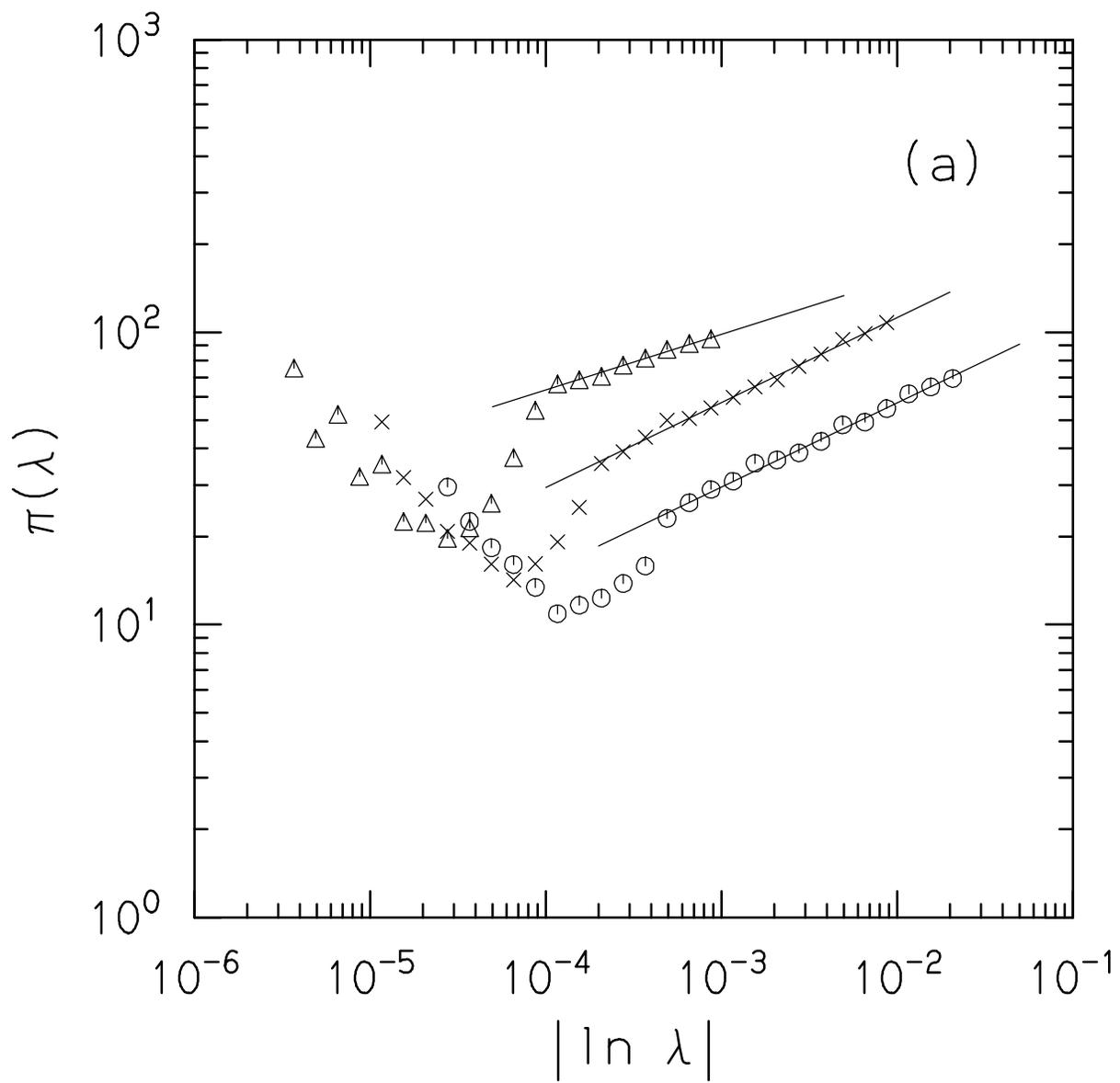
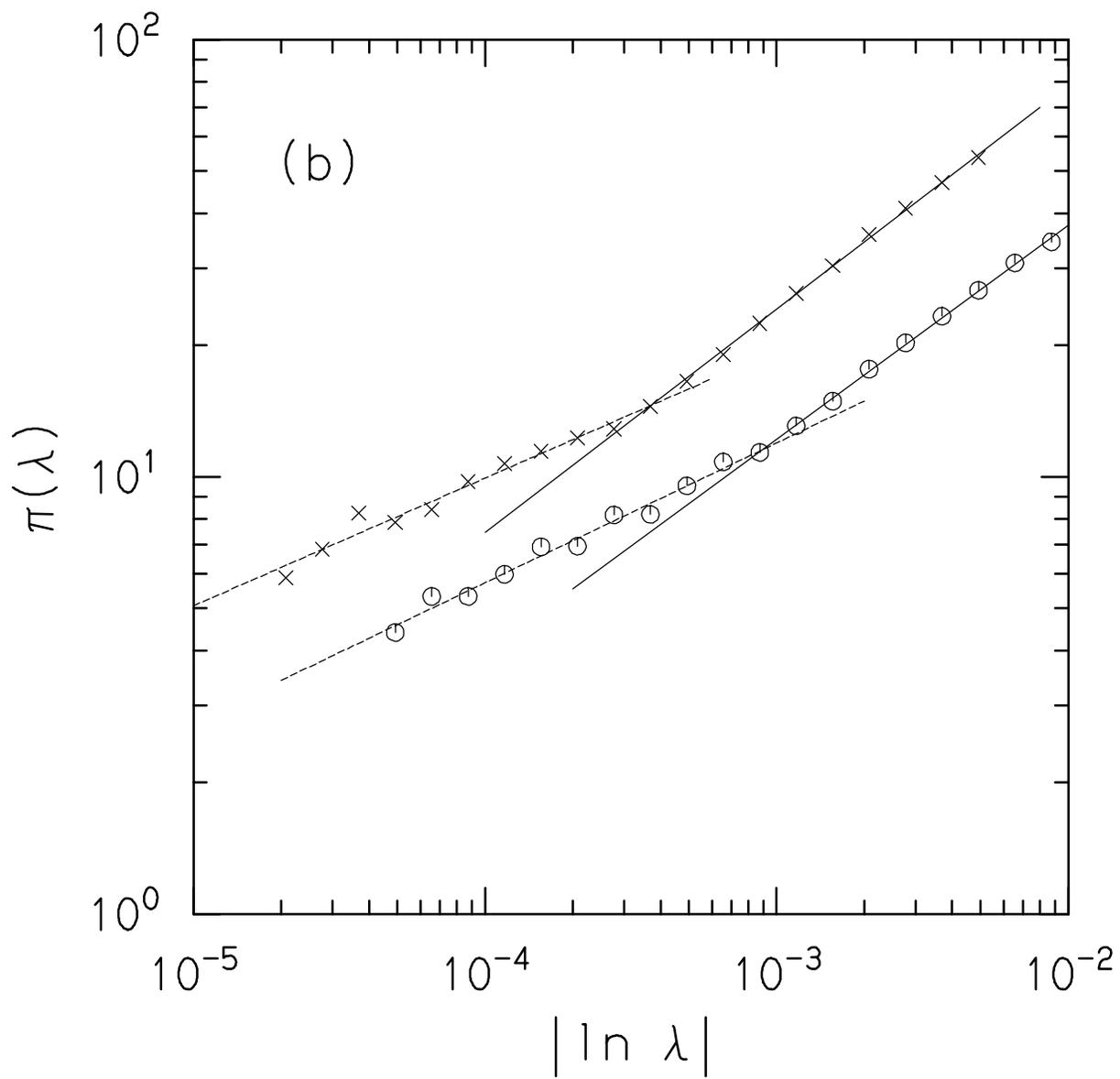
\bye